\documentstyle[times,11pt,epsf,fleqn,psfig]{article}    

\def\reference{\parskip 0pt\par\noindent\hangindent 0.5 truecm}
 
\def\cygx1{Cyg~X-1}  
\def\cirx1{Cir~X-1} 
\def\g321{G321.9$-$0.3}
\def\gx339{GX~339$-$4} 
\def\grs1915{GRS~1915+105} 
\def\gro1655{GRO~J1655$-$40}  

\newcommand{\solar}{\ifmmode{_{\mathord\odot}}\else%
  {$_{\mathord\odot}$}\fi}
\newcommand{\Msolar}{\ifmmode{\; {\rm M\solar}}\else%
  {${\; {\rm M\solar}}$}\fi}
\newcommand{\ergs}{\ifmmode{\;{\rm erg\,s^{-1}}}\else%
  ${\;{\rm erg\,s^{-1}}}$\fi}
\newbox\grsign \setbox\grsign=\hbox{$>$}
\newdimen\grdimen \grdimen=\ht\grsign
\newbox\laxbox \newbox\gaxbox
\setbox\gaxbox=\hbox{\raise.5ex\hbox{$>$}\llap
        {\lower.5ex\hbox{$\sim$}}}\ht1=\grdimen\dp1=0pt
\setbox\laxbox=\hbox{\raise.5ex\hbox{$<$}\llap
        {\lower.5ex\hbox{$\sim$}}}\ht2=\grdimen\dp2=0pt
\newcommand{\simlt}{\mathrel{\copy\laxbox}}
\newcommand{\simgt}{\mathrel{\copy\gaxbox}}

\begin{document}  
 
\parskip 0.725ex 
\baselineskip=1.04em
\renewcommand{\topfraction}{0.8}
\renewcommand{\bottomfraction}{0.5}

\title{Populations of X-ray binaries and  \\ 
  the dynamical history of their host galaxies}  

\author{Kinwah Wu \\ 
        Research Centre for Theoretical Astrophysics, 
        School of Physics, \\ 
        University of Sydney, NSW 2006, Australia, and \\ 
        Mullard Space Science Laboratory, University College London,  \\  
        Holmbury St Mary, Surrey RH5 6NT, United Kingdom  \\ 
        e-mail: kw@mssl.ucl.ac.uk } 

\maketitle   

\abstract{ 
The observed luminosity distributions of X-ray sources  
  indicate the presence of several populations of X-ray binaries 
  in the nearby galaxies.  
Each population has its formation and evolutionary history, 
  depending on the host environment. 
The features seen in the log~N($>$S) -- log~S curves 
  for different types of galaxies and for different galactic components   
  can be reproduced by a birth-death model, 
  in which the lifespans of the binaries   
  are inversely proportional to their X-ray brightness.    
Conversely, the dynamical history of a galaxy can be inferred 
  from the luminosity distributions of its X-ray binary populations. }  

\section{Introduction}  

X-ray binaries are powered by a compact star,  
  which may be a neutron star or a black hole, accreting material 
  from its companion.  
Systems having a massive OB companion star  
  are called high-mass X-ray binaries (HMXBs), 
  and systems with a low-mass companion star are known as    
  low-mass X-ray binaries (LMXBs).  
In HMXBs the compact star accretes via 
  capturing the stellar wind from its companion;   
  mass transfer in LMXBs occurs  
  when the companion star overflows its Roche lobe.   

X-ray binaries in the active states are luminous     
  and easily detected within the Galaxy.  
Many X-ray binaries were also found in nearby galaxies 
  by the {\it Einstein} and {\it ROSAT} X-ray satellites  
  (Fig.~1, see also e.g.\ Fabbiano 1995; Roberts \& Warwick 2000), and  
  recently, more are discovered by the {\it Chandra} X-ray observatory 
  (see Weisskopf et al.\ 2000 for the description of the observatory).  
It is now evident that galaxies similar to our own 
  (e.g.\ M31, Supper et al.\ 1997) 
  normally host hundreds of active X-ray binaries.   
The current sample of X-ray binaries in external galaxies 
  is sufficiently large  
  that not only is reliable population analysis 
  for an individual galaxy possible 
  but we can also study their formation and evolution 
  in different galactic environments.   

In this paper we present a simple birth-death model 
  and calculate the populations of X-ray binaries in external galaxies.  
We also use the model to identify the relevant processes 
  that give rise to the features 
  in the luminosity distributions of X-ray binaries  
  (such as the log~N($>$S) -- log~S curves) 
  in different types of galaxies and in different galactic components.  
Here, only the basic formulation is presented 
  and a few simple cases are shown as an illustration. 
Results of a more comprehensive study will be reported 
  in Wu et al.\ (2001).  

\section{X-ray binaries in a galaxy}    

\subsection{Accretion luminosity} 
  
The luminosity of an accreting compact star is given by 
\begin{equation}  
  L_{\rm x} \ \approx \  {GM {\dot M} \over R} 
    \ = \ 2.8\times 10^{37}~\biggl({M\over{{\rm M}_\odot} }\biggr) 
    \biggl({{\dot M}\over{10^{-8}{\rm M}_\odot{\rm yr}^{-1}}} \biggr) 
    \biggl({R\over{3\times 10^6{\rm cm}}}\biggr)^{-1}~{\rm erg~s}^{-1} \ , 
\end{equation}   
  where $G$ is the gravitational constant, 
  $M$ and $R$ are the mass and radius of the compact star, 
  and $\dot M$ is the mass accretion rate. 
Typically, neutron-star binaries have luminosities      
  $L_{\rm x} \sim 10^{37}$~erg~s$^{-1}$,   
  but the luminosities of binaries with accreting white dwarfs 
  would not greatly exceed $10^{33}$~erg~s$^{-1}$.   

\begin{figure} 
\begin{center}
\epsfxsize=7.5cm   \epsfbox{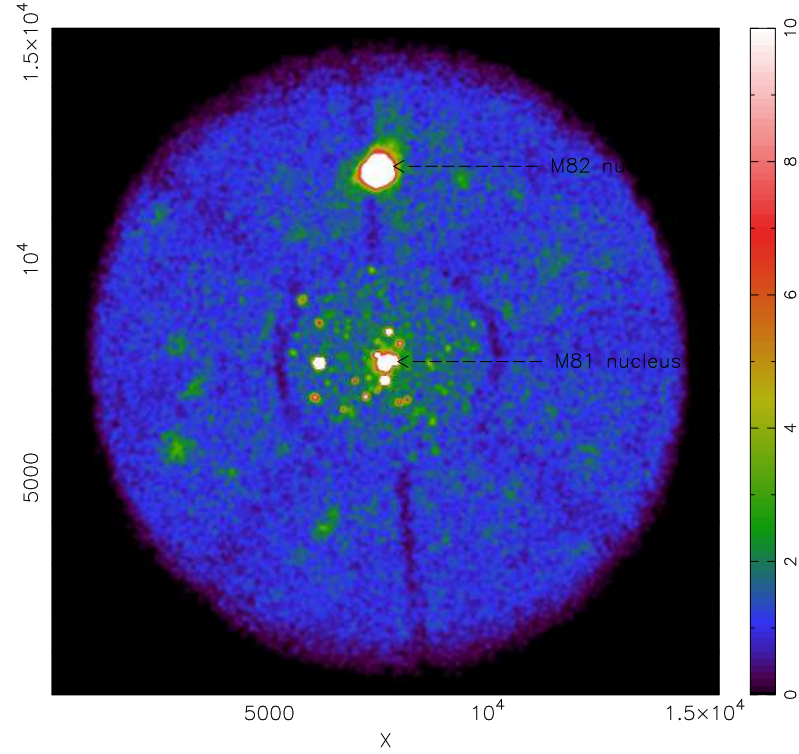}     
\end{center}      
\caption{\scriptsize 
    {\it ROSAT} PSPC image of the spiral galaxy M81 and  
       the starburst galaxy M82. 
    The brightest point sources in M81 can be easily distinguished 
       from the background.     }  
\label{fig.1}           
\end{figure}  
 
The expression for $L_{\rm x}$ above is not strictly applicable
  to black-hole binaries, 
  as black holes do not have a solid stellar surface that defines $R$. 
However, as accreting material attain speeds  
  close to the light speed, $c$,  
  when crossing the black-hole event horizon, 
  we may parametrise the accreting luminosity as     
\begin{equation}  
  L_{\rm x}\ \approx \ \eta_{\rm bh}~{\dot M c^2} 
   \ = \ 5.7\times 10^{37}~\biggl({\eta_{\rm bh} \over 0.1}\biggr) 
    \biggl({{\dot M}\over{10^{-8}{\rm M}_\odot{\rm yr}^{-1}}} \biggr) 
    ~{\rm erg~s}^{-1}  \ ,      
\end{equation} 
  where $\eta_{\rm bh}$ is the efficiency. 
If most of the X-rays are emitted from the inner accretion disk
  near the last stable orbit, 
  then $\eta_{\rm bh} \approx 0.1$ for a Schwarzschild black hole 
  and $\eta_{\rm bh} \approx 0.4$ for a maximally rotating Kerr black hole.  

The characteristic luminosity of neutron-star binaries 
  can also be expressed as  
  $ L_{\rm x} \approx \eta_{\rm ns}~{\dot M c^2}$,   
  because of the tight mass distribution of neutron stars.      
If we assume a canonical mass of 1.5~M$_\odot$ for the neutron stars, 
  then the value of the efficiency parameter $\eta_{\rm ns}$ is about 0.3.  
Hence, we can adopt    
  $L_{\rm x}\ \approx \ \eta~{\dot M c^2}$ (with $\eta \sim 0.1$)   
  for both neutron-star and black-hole binaries,  
  provided that the mass-transfer rates $\dot M$ 
  are not significantly above the Eddington limit.   

\subsection{Eddington limit} 
 
The infall of the material onto a compact star is opposed  
  by the accretion-generated radiative pressure, 
  and the accretion process is self-regulating. 
The limiting luminosity $L_{\rm ed}$, known as the Eddington luminosity, 
  for spherical accretion is given by   
\begin{equation}  
  L_{\rm ed} \ \approx \ 
      4\pi \biggl({G M m_{\rm p}c \over \sigma_{\rm T} }\biggr)  
    \ = \ 1.3\times 10^{38}~ 
       \biggl({M\over{{\rm M}_\odot} }\biggr) ~{\rm erg~s}^{-1} \ , 
\end{equation} 
  where $m_{\rm p}$ is the proton mass, 
  and $\sigma_{\rm T}$ is the Thomson cross section. 
The Eddington luminosity depends linearly  
  on the mass of the accreting star 
  and is independent of other parameters.  

As the average luminosity of an accretor would not be greater than    
  the Eddington luminosity, 
  the lower limit to the mass of the accretor can be constrained 
  by the observed luminosity.  
Thus, we can use this as a working criterion 
  to identify black-hole candidates. 
For instance, when we observe a source with X-ray luminosities 
  greatly exceeding  $2\times 10^{38}$~erg~s$^{-1}$,
  the Eddington luminosity 
  of a 1.5-M$_\odot$ compact star,   
  we may classify it as a (candidate) black-hole system 
  (if it is powered by accretion).   

\subsection{Evolution and expected lifespan}   

The duration of the active phase of an X-ray binary 
  is limited by the lifespan of the companion star, 
  and it is almost independent of the nature of the compact star. 
Massive OB stars have lifespans of $\sim 10^6-10^7$~yr, 
  implying that HMXBs powered by capturing the stellar wind 
  from massive OB stars   
  will not have a lifespan 
  significantly longer than a few million years. 
The presently active HMXBs must therefore have been formed 
  in very recent epochs.   
   
Low-mass stars have longer lifespans and evolutionary timescales 
  than high-mass stars,  
  and so LMXBs can remain X-ray active over a period longer than $10^7$~yr.   
Mass transfer in LMXBs is usually driven 
  by Roche-lobe overflow, 
  which is caused either by the expansion of their companion stars 
  when they evolve towards a red-giant phase,   
  or by orbital shrinkage, 
  when the binaries lose orbital angular momentum. 

For the first type of LMXBs,  
  their mass-transfer rate is roughly given by 
  $\langle{\dot M}\rangle \sim M_2/\tau_{\rm ev}$,  
  where $M_2$ is the mass of the companion star,  
  and $\tau_{\rm ev}$ is the evolutionary timescale. 
The nuclear-evolution timescale of a star 
  depends on its mass and evolutionary stage. 
The lifespan of F main-sequence stars is about $5 \times 10^9$~yr;  
  stars later than G type take more than $10^{10}$~yr
  to evolve away from the main-sequence stage.  
Clearly, nuclear evolution of main-sequence stars 
  of masses $\sim 1{\rm M}_\odot$ 
  is unable to sustain a persistent accretion luminosity 
  $\simgt 10^{37}$~erg~s$^{-1}$. 
Such accretion luminosities are possible 
  for LMXBs with evolved companion stars.       
The duration of the red-giant phase 
  is $\simlt 10^7$~yr for low mass stars (de Loore \& Doom 1992),  
  and hence, X-ray binaries with a red-giant companion are short-lived.    
However, galactic stars evolve into the red-giant stage at all epochs.  
X-ray binaries with red-giant companions are therefore formed continually 
  throughout the life time of a galaxy. 
 
For the second type of LMXBs, 
  orbital angular momentum can be extracted from the systems  
  by emitting gravitational radiation (Paczynski 1967) 
  or by rotational braking of the companion star 
  via a magnetic stellar wind (Verbunt \& Zwaan 1981).       
The timescale of orbital evolution driven by gravitational radiation is   
\begin{eqnarray}  
  t_{\rm gr} & = & {5\over {32(2\pi)^{8/3}}}{c^5 \over G^{5/3}} 
     { {(M_1 + M_2)^{1/3}}\over {M_1 M_2}} P^{8/3} \nonumber \\ 
     & = &  3.8\times 10^{11}\ (1+q)^{1/3} q^{-1} 
        \biggl({{M_1}\over {{\rm M}_\odot} } \biggr)^{-5/3} 
        \biggl({P \over {\rm 1~day} } \biggr)^{8/3} \ {\rm yr}   
\end{eqnarray}  
  (Landau \& Lifshitz 1971, see also Wu 1997), 
  where $P$ is the orbital period, 
  $M_1$ and $M_2$ are the masses 
  of the compact star and the companion star respectively, 
  and $q~(\equiv M_2/M_1)$ is the mass ratio.  
As the timescale of orbital shrinkage is long ($\gg 10^9$~yr), 
  the corresponding mass-transfer rate is not large enough  
  to account for the observed luminosity of the X-ray binaries.  
The orbital evolutionary timescale for magnetic braking is  
\begin{eqnarray}  
   t_{\rm mb} & = &  {{2 \times 10^{28}} \over {(2\pi)^{10/3}}}  
    {{f^2 G^{2/3}} \over {{\tilde k}^2}} 
    {{{M_\odot}^{4\gamma}}\over{{R_\odot}^4}} 
    {{M_1~P^{10/3}}\over{(M_1+M_2)^{1/3}M_2^{4\gamma}}} \ {\rm sec} \\ 
   & = & 4.4\times10^9 \ f^2 (1+q)^{-1/3} q^{-4\gamma} 
         \biggl({{\tilde k}^2 \over 0.1 }\biggr)^{-1}  
         \biggl({M_1 \over {{\rm M}_\odot}} \biggr)^{(2-12\gamma)/3} 
         \biggl({P\over{\rm 1~day}}\biggr)^{10/3} \ {\rm yr}   
\end{eqnarray}  
  (Verbunt \& Zwaan 1981; Wu 1997).  
The parameters ${\tilde k}$ and $f$ 
  depend on the stellar model (see Skumanich 1972), 
  and the parameter $\gamma$ is determined by the mass-radius relation  
  ($R_2 \propto M_2^\gamma$, where $R_2$ is the radius of the companion).  
For low-mass stars, ${\tilde k}^2 \approx 0.1$, $f \sim 1$ 
  and $\gamma \approx 1$.  
The mass-transfer rate is higher than 
  that in the case of gravitational radiation, 
  and the active phase of these X-ray binaries 
  usually lasts $\simgt 10^8$~yr.    
  
\section{Populations of X-ray binaries}  

In a simplistic point of view 
  the populations of X-ray binaries in a galaxy observed at an epoch $t$ 
  are determined by the birth rates and the lifespans of the binaries.  
Assuming other factors are unimportant  
  (for example, the two-body and three-body capture process 
  in globular clusters, see Johnston \& Verbunt 1996),  
  we can construct a birth-death model 
  and calculate the populations and luminosity distributions  
  of X-ray binaries in a galaxy 
  or in a particular galactic component.  

The basic formulation is as follows. 
Let the number of the X-ray sources 
  at the luminosity range $(L,L+dL)$ in a galaxy at time $t$ 
  be $n(L,t)dL$ and 
  the characteristic lifespan of these sources be $\tau$. 
The evolution of an X-ray binary population is governed by  
\begin{eqnarray}  
    {d \over dt} n(L,t) & = & -k~n(L,t) + f(L,t) \ ,   
\end{eqnarray}  
  where $k~n(L,T)$ is the death rate, 
  the $f(L,t)$ is the birth rate,  
  and $k^{-1}$ ($\equiv \tau$) is the characteristic lifespan 
  of the sources.      
The number of sources with a luminosity brighter than $L$ is simply 
\begin{eqnarray}  
  N(>L,t) & = & \int_L^\infty dL ~ n(L,t) \ .  
\end{eqnarray}    
When the functional forms of $k$ and $f(L,t)$ are specified, 
  Equations (7) and (8) can be solved easily.    

\subsection{Impulsive birth}   

X-ray binaries are born (i.e.\ become X-ray active)  
  when mass transfer from the companion to the compact star starts;  
  and they die when the X-ray active phase ends. 
Suppose that there are two channels to create X-ray binaries in a galaxy:  
  (i) systems that are formed continuously 
  and the mass transfer of these systems is sustained  
  by nuclear evolution of the companion star or 
  by the orbital evolution of the binary, and   
  (ii) systems that are born in a recent star-formation episode.      
The birth rate function $f(L,t)$ thus consists of  
  a steady (continual birth) component 
  and an impulsive (starburst) component:    
\begin{eqnarray} 
  f(L,t) & =  & f_o(L) \big[ 1 + a \delta(t-t_a) \big]  \ ,   
\end{eqnarray}    
  where $f_o(L)$ is the birth rate of the steady component,  
  $a$ is the ratio of the strength of the impulsive component 
  to the strength of the steady component,  
  $\delta(t-t_a)$ is the Dirac $\delta$-function, and   
  $t_a$ is the time at which the starburst occurred. 
  
The effective duration of the active phase of an X-ray binary     
  is determined by the mass of the companion star 
  and the mass-transfer rate, and 
  it is approximately given by $M_2/\langle {\dot M}\rangle$. 
As $L_{\rm x} \propto {\dot M}$,  
  the characteristic lifespan of an X-ray binary $k^{-1}$ 
  can be parametrised as $k =  \beta L$, 
  where $\beta^{-1} = {\tilde \eta} \langle M_2\rangle c^2$ and 
  $\langle M_2 \rangle$ is the mean mass of the companion star, 
  and ${\tilde \eta}$ is a parameter 
  which has the same order of magnitude 
  as the parameters $\eta_{\rm bh}$ and $\eta_{\rm ns}$ in \S 2.    
The fact that the characteristic lifespan of binaries 
  with luminosities in the range ($L, L+dL$) 
  is not explicitly time-dependent  
  allows us to integrate Equation (7) directly, 
  yielding          
\begin{eqnarray}  
  n(L,t) & = & n_o(L)~e^{-\beta L t} + {{f_o(L)} \over {\beta L}}
        \biggl( 1-e^{-\beta L t} \biggr) 
         + {a f_o(L)}   
        \theta (t - t_a) e^{-\beta L(t-t_a)} \ ,   
\end{eqnarray}   
  where $\theta (t-t_a)$ is the Heaviside unit step function, 
  and $n_o(L) \equiv n(L,0)$ is the initial population.   

If the initial population and the birth rate are power-laws of $L$, i.e.\ 
  $n_o = n_{o*} (L/L_*)^{-\alpha_1}$ and $f_o = f_{o*} (L/L_*)^{-\alpha_2}$ 
  (with power-law indices $\alpha_1$ and $\alpha_2$ respectively), then   
\begin{eqnarray}  
  N(>L,t) & = &  n_{o*} L_* (\beta L_* t)^{\alpha_1-1} 
    \Gamma \big(1-\alpha_1, \beta L t \big)  \nonumber \\ 
    & &  +\   {{f_{o*}} \over \beta} (\beta L_* t)^{\alpha_2}~    
      \biggl[{1 \over {\alpha_2}} (\beta L t)^{-\alpha_2}
          \biggl( 1 - e^{-\beta L t} \biggr)  
         + {1 \over {\alpha_2}}  
      \Gamma\big(1 -\alpha_2, \beta Lt \big) \biggr] \nonumber \\ 
    & &  +\ {a f_{o*}} L_*[\beta L_* (t-t_a)]^{\alpha_2-1}~ 
         \Gamma\big(1-\alpha_2, \beta L (t-t_a) \big)  \theta (t-t_a) \ , 
\end{eqnarray} 
  where $L_*$ is a lower cut-off luminosity, 
  and $\Gamma (\alpha , x)$ is the incomplete gamma function.    
In the limits of $f_{o*} \rightarrow 0$  
  (i.e.\ all the sources are primordial)   
  the X-ray binary population has a distribution 
  similar to the Schechter (1976) analytic luminosity function 
  for galaxies in clusters.  
The same distribution is also obtained 
  when $(f_{o*} /\beta L_*) \gg n_{o*}$ and $a \beta L_* \gg 1$
  (i.e.\ most of the binaries were born in a recent starburst).    
The difference between the Schechter (1976) luminosity function 
  and the distribution that we obtain here is that 
  in the latter case the luminosity break is time dependent --- 
  it is caused by the aging of the source population  
  when it is not replenished.    

\begin{figure}
\begin{center}         
\epsfxsize=7.5cm   \epsfbox{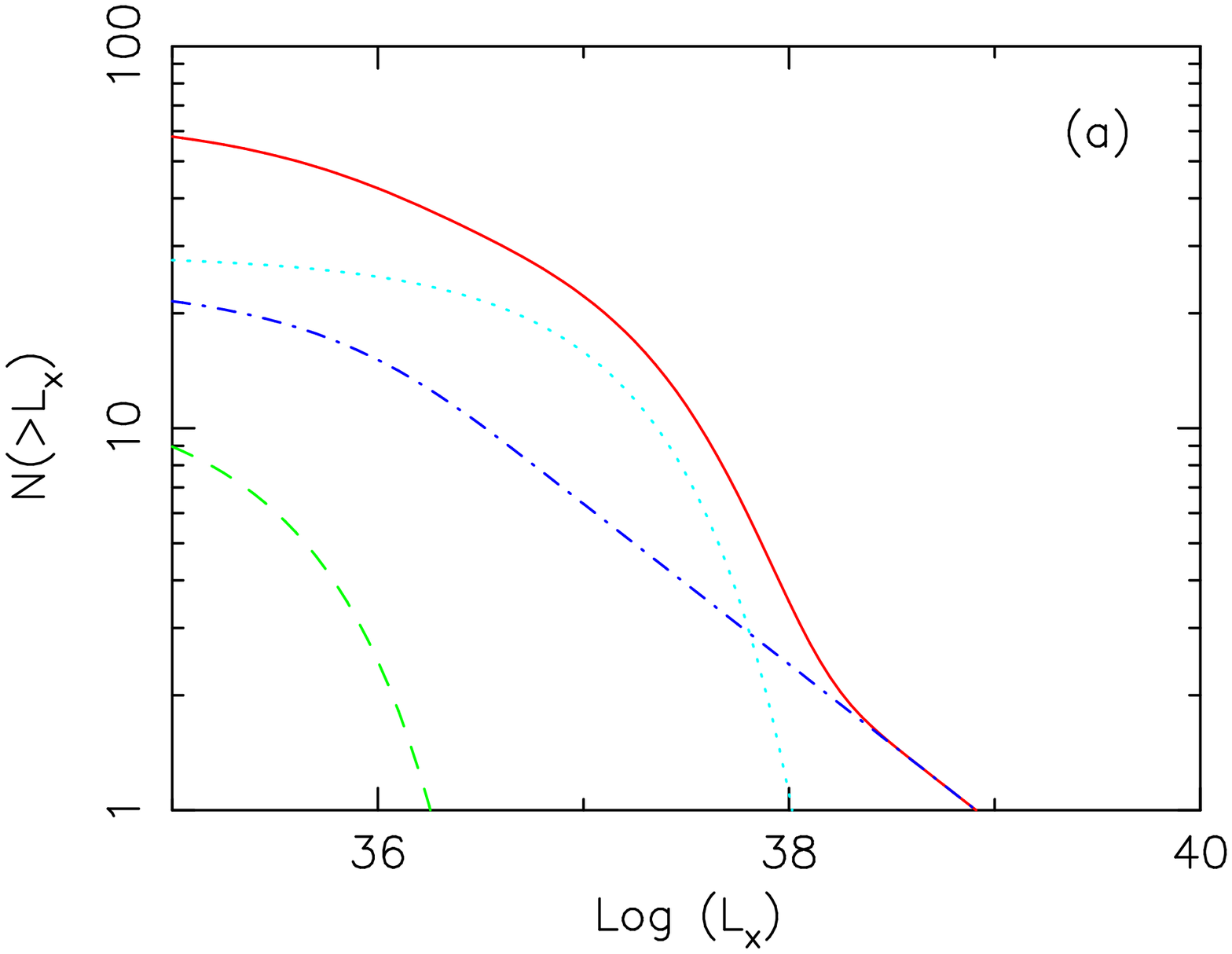}    
\epsfxsize=7.5cm   \epsfbox{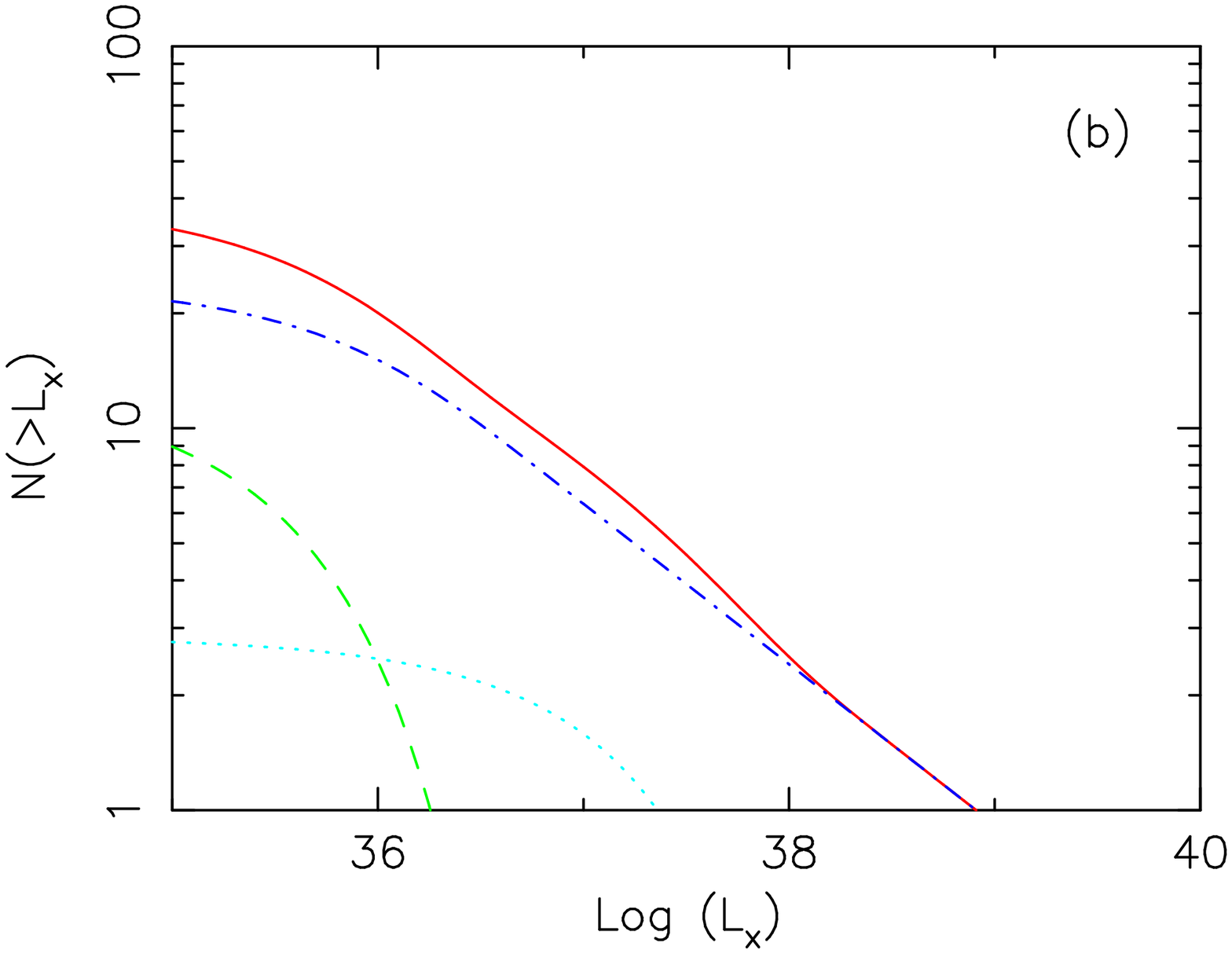}  
\epsfxsize=7.5cm   \epsfbox{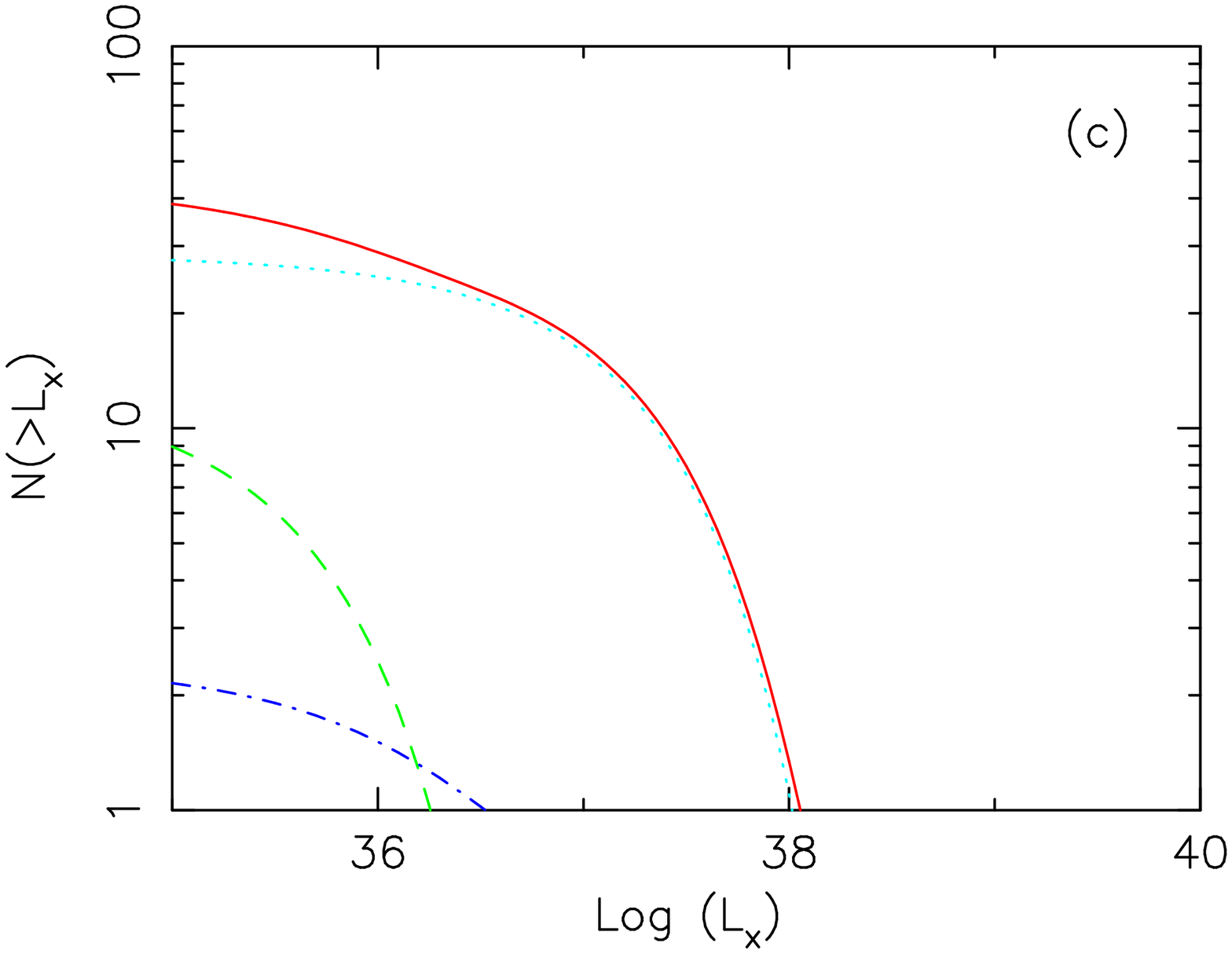}  
\epsfxsize=7.5cm   \epsfbox{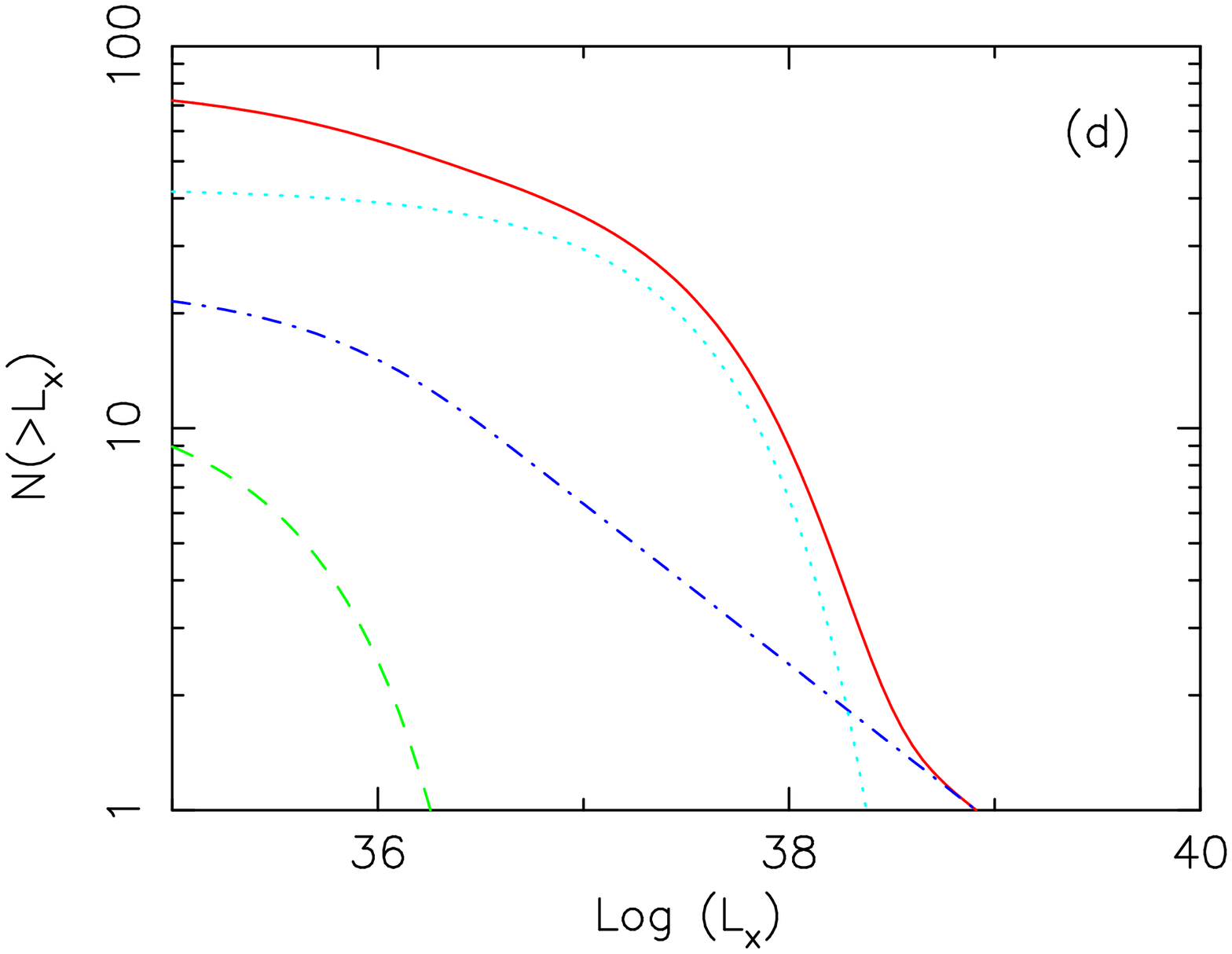}   
\end{center}      
\caption{\scriptsize 
  Luminosity distributions of X-ray binaries 
     in model galaxies for the impulsive-birth model.  
  In the calculations we assume an age of 15~Gyr 
     for the Universe.    
  We also define two parameters  
     $\lambda_c \equiv (f_{o*}/n_{o*}\beta L_*)$ 
     and  $\lambda_a \equiv (a f_{o*}/n_{o*})$ 
     to specify the relative strength 
     of the continuous and the impulsive component respectively. 
  The power-law indices $\alpha_1$ and $\alpha_2$     
       are 0.5 and 0.42 respectively.    
  (a): The starburst epoch occurred at 400 million years ago, 
       i.e.\ $t_{\rm H}- t_a = 0.4$~Gyr.     
     The values for $\lambda_c$ and $\lambda_a$ are 
       1.0 and 0.3 respectively. 
     The primordial component is represented by the dashed line; 
       the continuous component, by the dot-dashed line; and 
       the impulsive component, by the dotted line. 
     The total population is represented by the solid line.              
     The normalisation is chosen such that the total source counts 
       are similar to the counts of the bulge sources 
       in M81 observed by {\it Chandra}  
       (see Tennant et al.\ 2001).    
  (b): Same as (a) except $\lambda_a = 0.03$.  
  (c): Same as (a) except $\lambda_c = 0.1$.  
  (d): Same as (a) except for $t_{\rm H}-t_a =0.2$~Gyr.  }  
\label{fig.2}             
\end{figure}             
  
\subsection{Cyclic generation}   

In reality, star formation may not be a single event. 
For example, when two galaxies are orbiting around each other,   
  periodic starbursts are induced by orbital interaction.
In this case, 
  the birth term $f(L,t)$ has a periodic (starburst) component 
  superimposed on a steady background: 
\begin{eqnarray} 
  f(L,t) & =  & f_o(L) \big[ 1 + b \cos \omega t \big] \ ,     
\end{eqnarray}    
  where $2\pi /\omega$ is the characteristic timescale of the cycle.   
It follows that the population of X-ray binaries at time $t$ is    
\begin{eqnarray}   
  n(L,t) & = & n_o(L)~e^{-\beta L t} + {{f_o(L)} \over {\beta L}}
        \biggl( 1-e^{-\beta L t} \biggr)  \nonumber \\ 
     & &   + \ {{b f_o(L)} \over {\beta L}}   \biggl[ 
         \biggl( 1-e^{-\beta L t} \biggr) 
         +  {{(\beta L)^2} \over{(\beta L)^2 + \omega^2}}    
     \biggl( 1 - \cos \omega t - 
        {\omega \over {\beta L}} \sin \omega t \biggr) \biggr]  \ ,     
\end{eqnarray}  
   and the number of systems brighter than $L$ is 
\begin{eqnarray}  
    N(>L,t) & = &  n_{o*} L_* (\beta L_* t)^{\alpha_1-1} 
    \Gamma \big(1-\alpha_1, \beta L t \big)  \nonumber \\ 
    & &  +\   {{f_{o*}} \over \beta}  (\beta L_* t)^{\alpha_2}~    
      \biggl[{1 \over {\alpha_2}} (\beta L t)^{-\alpha_2}  
      \biggl(1 - e^{-\beta L t}  \biggr)  
         - \Gamma\big(1 -\alpha_2, \beta Lt \big) \biggr] \nonumber \\ 
    & &  +\ {{bf_{o*}} \over \beta}  (\beta L_* t)^{\alpha_2}~    
     \bigg\{ \biggl[{1 \over {\alpha_2}} (\beta L t)^{-\alpha_2}  
        \biggl(1 - e^{-\beta L t}  \biggr) 
         + \Gamma\big(1 -\alpha_2, \beta Lt \big) \biggr]  \nonumber \\ 
   & & \hspace*{0.5cm}  
     +\  \biggl[~ I\big(\alpha_2 -1; \omega t, \beta L t\big) 
          ( 1- \cos \omega t )  - \ 
   I\big(\alpha_2; \omega t, \beta L t\big)(\omega t)\sin \omega t ~  
    \biggr] \bigg\}  \  .  
\end{eqnarray} 
The integral $I(\alpha; \omega t , \beta Lt)$ above is defined as   
\begin{eqnarray}  
   I\big(\alpha; \omega t, \beta L t \big) & \equiv & 
     \int^\infty_{\beta L t} dx ~ 
      {{x^{-\alpha}} \over {x^2 + (\omega t)^2}}  \  ,  
\end{eqnarray}   
   and it has exact analytic form for integer $\alpha$.   
The periodic component may  
  introduce ripple-like features in the luminosity function, 
  hence makes the log~N($>$S) -- log~S curves 
  deviate from a simple or a broken power law.   

\subsection{Other issues}   

\subsubsection{Neutron-star binaries} 

The model above has not accounted for the fact 
  that the persistent luminosity of neutron-star binaries 
  cannot exceed the Eddington limit of a 1.5-M$_\odot$ accretor. 
The luminosities of black-hole binaries are less restrictive, 
  as there is no practical mass limit for black holes.  
A more proper treatment requires the populations 
  of neutron-star and black-hole binaries 
  be calculated separately. 
Although Equation (7) is applicable for both classes of objects, 
  the corresponding expressions for $N(>L)$ are not identical.  
For a population of X-ray  binaries with a cutoff luminosity at $L_{\rm ed}$, 
  $N(>L)$ is given by  
\begin{eqnarray} 
   N(>L,t) & = & \int_L^{L_{\rm ed}} dL ~ n(L,t) \ .   
\end{eqnarray}  
The tight mass distribution of neutron stars 
  together with the presence of a luminosity limit 
  will produce a spike at $L_{\rm ed}$   
  in the (differential) luminosity function $n(L)$, 
  and hence a cut-off in the observed log~N($>$S) -- log~S curve. 

The strength of the spike in $n(L)$ is time-dependent.   
If all the neutron-star binaries in the galaxy 
  were born at the same episode, 
  because the systems with high mass-transfer rates 
  cease to be active first 
  and then those with lower mass-transfer rates, 
  the spike will gradually decrease, 
  on a timescale    
\begin{eqnarray} 
  t_{\rm sp} & \simlt & {{\langle M_2 \rangle}\over{{\dot M}_{\rm ed}}} \\ 
   & \approx 
    & 1.3 \times 10^8~{M_2\over M_{\rm ns}}~{\rm yr} \ , 
\end{eqnarray} 
  where ${\dot M}_{\rm ed}$ and $M_{\rm ns}$
  is the Eddington mass-accretion rate and 
  the mass of neutron stars respectively.  
For galaxies with star-formation activity 
  occurred more than $10^8$~yr ago,   
  the spike has already degraded so substantially 
  that it may not be detected easily (Wu et al.\ 2001).  
The spike is, however, prominent 
  for galaxies with violent starburst activity   
  in the near past ($\simlt 10^7$~yr), 
  or if the neutron-star binaries are formed continuously.  

\subsubsection{Transient sources} 

The total duration of the X-ray active phases of transients 
  is only some fraction of their lifespans. 
The parameter $k$ is therefore not exactly equal to $\dot M/M_2$, 
  but is proportional to it. 
As a first approximation, 
  we may use an effective factor $\xi$ 
  to specify the fractional time   
  at which the source is X-ray active, 
  i.e.\ $k = \beta L /\xi $.  
If $\xi$ does not depend on $L$ explicitly, 
  the solution to Equation (7) preserves its form.  
Otherwise, the solution  
  and hence the luminosity functions are modified.  

It is worth noting that   
  in addition to different fractional time $\xi$
  for their X-ray active phases, 
  the transients and persistent sources 
  have different discovery probabilities.  
This could complicate the calculations 
  of the source populations and luminosity functions. 
Here, we have omitted the transient effects for simplicity, 
  but one must bear in mind that the effects might be important.    
  
\section{X-ray sources in external galaxies}  

Since 1999, {\it Chandra} has obtained high-resolution images 
  of a number of nearby galaxies.   
These include the ellipticals   
  M84 (NGC~4374) (Finoguenov \& Jones 2001)    
  NGC~4967 (Sarazin, Irwin \& Bregman 2000), 
  and Cen A (Kraft et al.\ 2000), 
  the lenticulars  NGC~1553 (Blanton, Sarazin \& Irwin 2001) 
  and NGC~1291 (Irwin, Bregman \& Sarazin 2001),  
  the spirals M31 (NGC~224) (Garcia et al.\ 2000),   
  M81 (NGC~3031) (Tennant et al.\ 2001), 
  M101 (NGC~5457) (Pence et al.\ 2001) and  
  the Circinus galaxy (Sambruna et al.\ 2001),    
  and the irregular starburst galaxy M82 (NGC~3034) 
  (Matsumoto et al.\ 2001), and 
  the mergers NGC~4038/4039 (Fabbiano, Zezas \& Murray 2001).     
The {\it Chandra} observations show   
  more than a hundred bright X-ray sources    
  (with $L_{\rm x} > 10^{36}$~erg~s$^{-1}$)   
  in the field each of these galaxies.   
While some of the bright sources are foreground stars and background AGN, 
  the majority are probably X-ray binaries within the galaxies.    
The observations also indicate that 
  the brightest sources in the early-type galaxies   
  are generally brighter than their counterparts 
  in the spiral galaxies, if the galactic nuclei are excluded.   

\begin{figure} 
\begin{center} 
\epsfxsize=7.5cm   \epsfbox{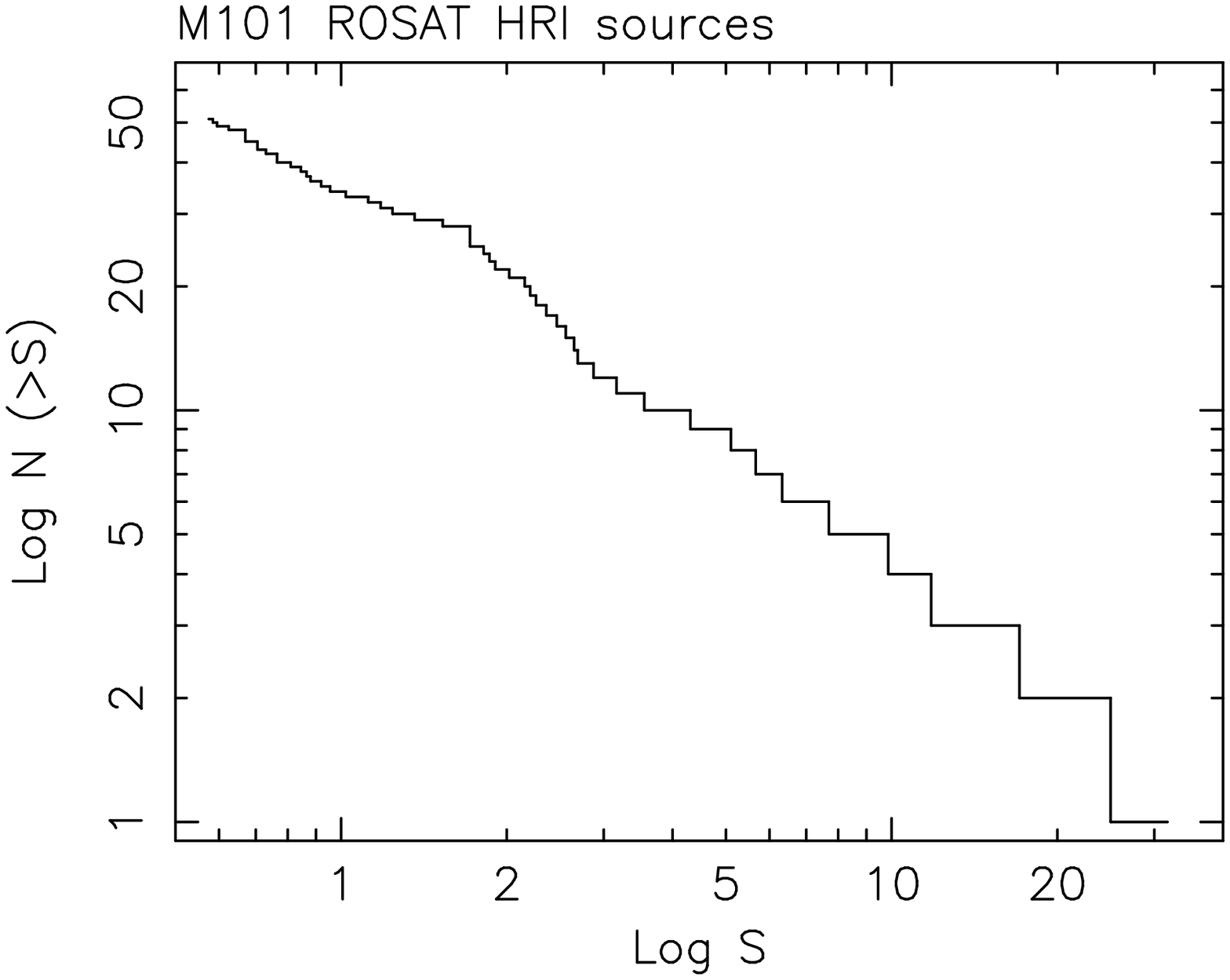} 
\epsfxsize=7.5cm   \epsfbox{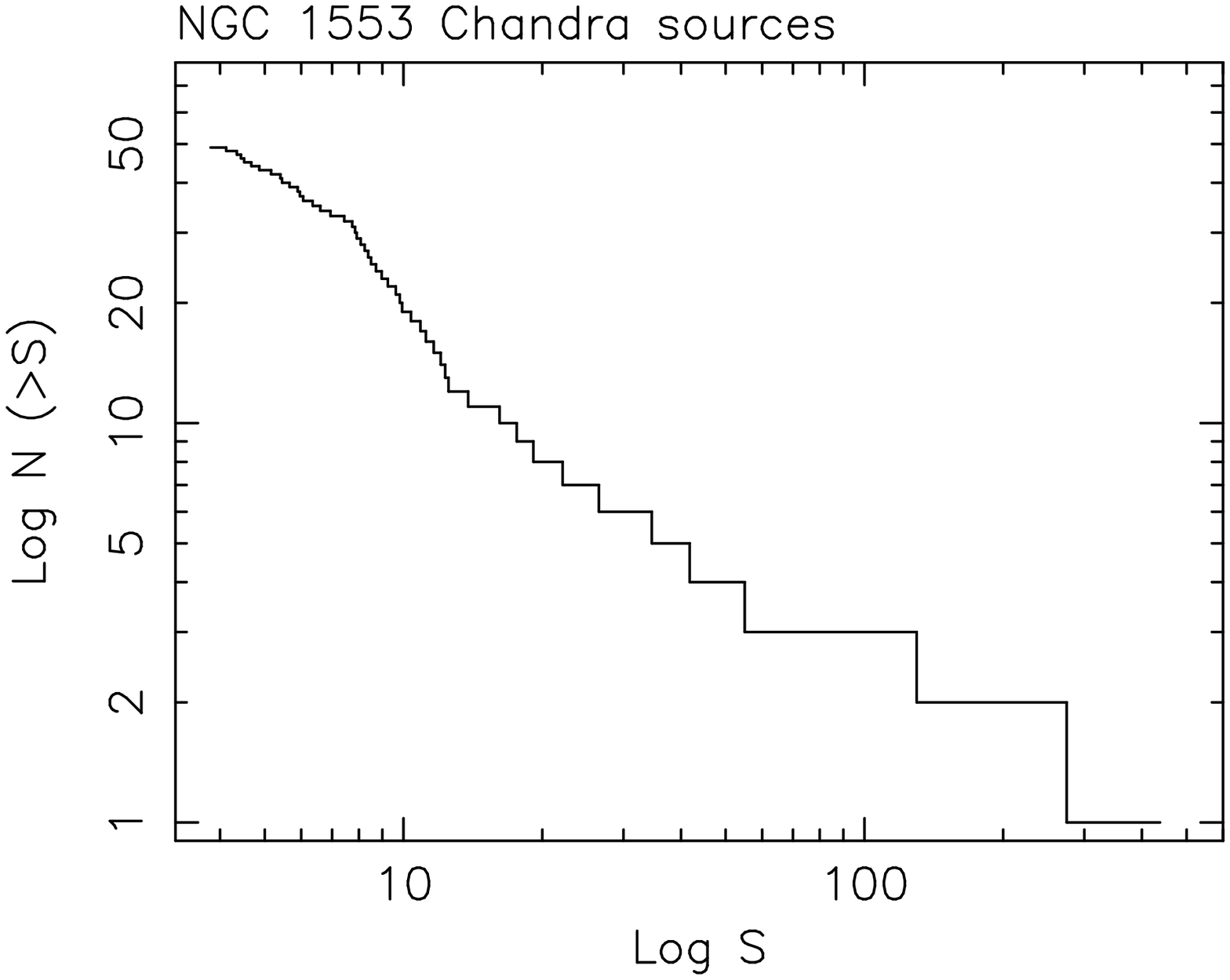}   
\end{center} 
\caption{\scriptsize  
  (Left) The log~N($>$S) -- log~S curve of the {\it ROSAT} HRI sources 
     in M101 (NGC~5457). 
  The data are taken from Table 3 in Wang, Immler \& Pietsch (1999). 
  The unit of S is counts per kilo-second. 
  For an assumed distance of 7.7~Mpc to the galaxy, 
     1 unit corresponds roughly 
     to a luminosity of $2.8\times 10^{38}$~erg~$^{-1}$ at 0.5--2.0~keV  
    (1 cts ks$^{-1}$ is approximately $4.0\times 10^{-14}$~erg~s$^{=1}$).  
  (Right) The log~N($>$S) -- log~S curve of the {\it Chandra} sources 
     in NGC~1553. 
  The data are taken from Table 1 in Blanton, Sarazin \& Irwin (2000). 
  The unit of S is counts per 10 kilo-second.  
  The conversion between count rate and the unabsorbed X-ray luminosity 
     is $4.91 \times 10^{41}$~erg~cts$^{-1}$ at the 0.3--10~keV 
     (see Blanton, Sarazin \& Irwin 2000 for details).   }  
\label{fig.3}
\end{figure} 
 
\subsection {log~N($>$S) -- log~S curve}   
 
The log~N($>$S) -- log~S curves 
  of the {\it Einstein} and {\it ROSAT} sources in spiral galaxies 
  are in general adequately fitted  
  by a power law (see Fabbiano 1995). 
However, for the galaxies with a sufficiently large number 
  of sources discovered ($\simgt$ a few tens), 
  the log~N($>$S) -- log~S curves are found to deviate significantly 
  from a single power law, e.g.\ M31 (see Supper et al.\ 1997) 
  and M101 (Fig.~3, left panel; see also Wang, Immler \& Pietsch 1999).  
A broken power law is often required  
  in order to fit the log~N($>$S) -- log~S curves.    
The break $L_{\rm c}$ is located at  
  $\sim {\rm a\ few} \times 10^{37}$~erg~s$^{-1}$ 
  (see e.g.\ M31, Supper et al.\ 1997; Shirey et al.\ 2001).   
 
The {\it Chandra} observations confirm  
  the presence of the broken power-law 
  in the log~N($>$S) -- log~S curves of X-ray sources 
  in spiral galaxies (e.g. M81, Tennant et al.\ 2001). 
They also verify that the broken power law is present   
  in the log~N($>$S) -- log~S curves for early-type galaxies,   
  e.g.\ NGC~4697 (Sarazin, Irwin \& Bregman 2000), 
  NGC~1553 (Fig.~3, right panel) 
  and M84 (A. Finoguenov, private communication).  
The fact that a luminosity break 
  is found in both the cases of early- and late-type galaxies
  suggests that the break may be universal. 
A possible explanation is that the luminosity break is caused 
  by a population of neutron-star binaries 
  which have super-Eddington mass-transfer rates.  
These neutron-binaries would have X-ray luminosities 
  roughly about $2\times 10^{38}$~erg~s$^{-1}$, 
  the Eddington luminosity of the 1.5-M$_\odot$ accretors.  
If the location of the luminosity break is the same 
  for all galaxies, it can be used as a distance indicator     
  (Sarazin, Irwin \& Bregman 2000). 
 
However, when the X-ray sources 
  in the disk and in the bulge of M81  
  are considered separately,   
  very different log~N($>$S) -- log~S curves 
  are obtained (Fig.~4, Tennant et al.\ 2001).   
The luminosity break is obvious   
  in the log~N($>$S) -- log~S curve of the bulge sources, 
  but the log~N($>$S) -- log~S curve of the disk sources 
  is a single power law.  
Interestingly, the power-law slope of the disk sources 
  is similar to that of the bulge sources  
  below the luminosity break.       
 
\begin{figure}
\begin{center}
\epsfxsize=10cm   
\epsfbox{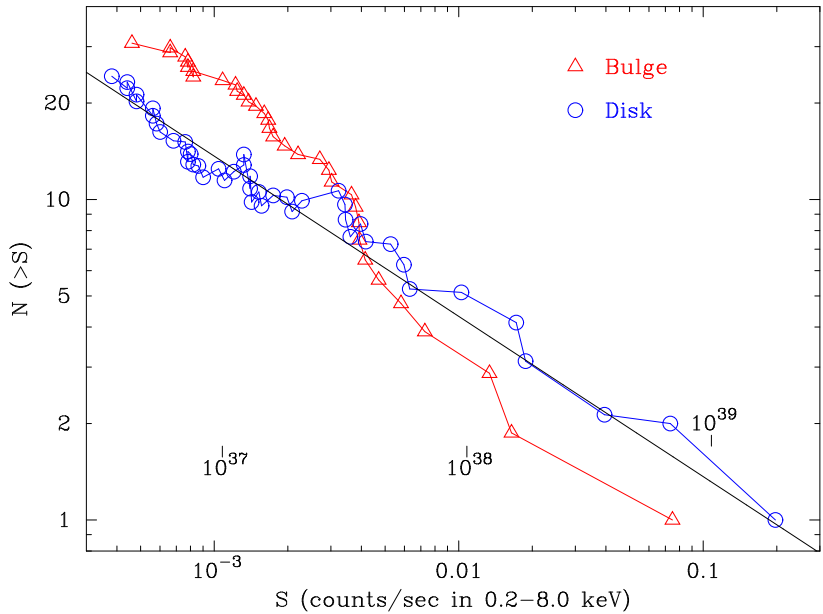}  
\end{center}   
\caption{\scriptsize 
  The log~N($>$S) -- log~S curves of the disk and bulge sources of M81 
     observed by {\it Chandra}.  
  The sources are background subtracted with 
     the log~N($>$S) -- log~S distribution of the background 
     taken from {\it Chandra} calibration observation of CRSS~J0030.5+2618. 
  Adopted from Tennant et al.\ (2001).}  
\label{fig.4} 
\end{figure}
  
\subsection{Dynamical evolution of the host galaxy}  

The absence of the break in the log~N($>$S) -- log~S curve 
  of the disk sources in M81  
  is inconsistent with the proposition that 
  the luminosity break is universal for all galaxies. 
The locations of the breaks 
  at different luminosities for early- and late-type galaxies 
  (cf.\ M81 and NGC~4697, 
  see Tennant et al.\ 2001 
  and Sarazin, Irwin \& Bregman 2000 respectively) 
  suggest that the natures of the breaks 
  are different for galaxies with different Hubble types.  
We propose that for some galaxies 
  the luminosity break is determined 
  by the formation and evolutionary history 
  of the populations of X-ray binaries in the galaxy. 
The different morphologies for the log~N($>$S) -- log~S curves 
  for the disk and bulge sources in the spiral galaxies  
  reflect the different star-formation histories 
  for the galactic components 
  and the different evolutionary paths 
  of their X-ray binary populations. 

We interpret the absence of the luminosity break 
  in the log~N($>$S) -- log~S curve 
  for disk sources in the spiral galaxy M81  
  as being a consequence of 
  a continuous, smooth star-formation process. 
In normal spiral galaxies, 
  star formation is triggered by compression of gas 
  when the density wave sweeps across the galactic disk. 
The star-formation process is continuous 
  and is relatively smooth 
  (in comparison with the environment 
  in starburst galaxies and mergers).  
As the X-ray binaries are formed continually 
  in the spiral arms,  
  the bright sources are dominated by young HMXBs 
  with substantial proportion of black-hole binaries. 
The disk sources may also consist 
  of two types of older LMXBs:  
  one with their companion stars   
  evolving into the red-giant phase, 
  another one 
  with their orbital separation sufficiently shrunk 
  that the systems become semi-detached.   
Many of these LMXBs may contain accreting neutron stars. 
However, if the young HMXBs dominate the total population,  
  there will not be a prominent luminosity break 
  in the log~N($>$S) -- log~S curve. 
 
Formation of X-ray binaries in a galactic component    
  is continuous if the galaxy is undisturbed. 
In this case, the population is dominated by LMXBs 
  which are formed via processes 
  the same as those of the LMXBs in the galactic disk. 
If the bulge is disturbed, 
  e.g.\ by tidal interaction with another galaxy,  
  a starburst is triggered by a sudden infall of gas,  
  and a new population of stars 
  as well as X-ray binaries are formed. 
As shown in \S 3.1, such an impulsive formation 
  can produce a prominent luminosity break 
  in the log~N($>$S) -- log~S curve, 
  and the location of the break. 
When the population of the X-ray binaries ages, 
  the break migrates to low luminosities, 
  thus marking the look-back time 
  of the catastrophic star-forming episode.  

Interestingly, the luminosity break 
  in the log~N($>$S) -- log~S curve 
  for the sources in the ellipticals (e.g.\ NGC~4697)  
  seems to occur at a higher luminosity than 
  that for the bulge sources in the spirals (e.g.\ M31 and M81)  
  (A.~F.~Tennant and D.~A.~Swartz, private communication). 
If luminosity breaks are created solely 
  by the aging of X-ray binary populations, 
  the break seen in the log~N($>$S) -- log~S curves 
  of the {\it Chandra} sources in the early-type galaxies 
  indicate that these galaxies had experienced 
  some, possibly violent, star-formation activity 
  in the very near past.  

However, this interpretation must be taken with caution. 
As pointed out in \S 3.3.1, 
  the tight mass distribution of neutron stars 
  together with the presence of an Eddington limit 
  for each accretor could produce a spike in $n(L)$ 
  and hence a luminosity cut off in $N(>L)$. 
A substantial proportion of neutron-star binaries 
  in the X-ray binary population  
  can also cause a luminosity break 
  in the log~N($>$S) -- log~S curve. 
The break caused by a population of neutron-star binaries 
  with massive companions is expected to be located 
  at $L_{\rm x} \approx 2 \times 10^{38}$~erg~s$^{-1}$, 
  the Eddington luminosity of a 1.5-M$_\odot$ accretor.  
Its strength will decrease on a timescale $\simlt 10^8$~yr, 
  if there is no replenishment of the systems. 
As the luminosity breaks  
  seen in the log~N($>$S) -- log~S curves 
  of the sources in the bulge of the spirals  
  scatter over a range of luminosities, 
  they are unlikely to be all due 
  to a population of accreting neutron stars.   
It is, however, possible that the breaks  
  in the log~N($>$S) -- log~S curves of the early-type galaxies 
  are caused by a population of neutron-star binaries 
  with high mass-transfer rates. 
  
\subsection{Super-Eddington sources}   

Although super-Eddington sources seem 
  to be more abundant in the early-type galaxies,  
  given the different stellar contents in the galaxies, 
  the excess of bright sources 
  in the early-type galaxies is yet to be confirmed. 
The finding that the brightest sources in the spirals 
  tend to be located in the disk 
  (e.g.\ in M81, Tennant et al.\ 2001) 
  implies that the disk sources are probably young systems. 
This also implies that the brightest sources 
  in the spiral galaxies 
  and the brightest sources in the early-type galaxies 
  may not form in the same way. 
The population of bright sources 
  in the arms of the spiral galaxies are dominated by HMXBs, 
  in which the mass donors are young massive OB stars 
  with strong winds;  
  while the population of bright sources 
  in the ellipticals and lenticulars are LMXBs, 
  with a red-giant companion overflowing its Roche lobe. 

Despite the conventional scenario 
  that X-ray sources in galaxies are mostly X-ray binaries, 
  the super-Eddington sources 
  may be an inhomogeneous class of exotic objects:  
  hypernovae, supernovae, supersoft sources, 
  intermediate-mass black holes, 
  large clusters of massive OB stars, .... etc. 
These sources have not been included in our calculations presented above, 
  and hence the results that we obtain 
  may not truly reflect the real populations of X-ray sources 
  in the galaxies. 
However, in order to take into account these objects, 
  one must have some working models 
  for their formation and evolution, 
  which are unfortunately not available at present. 
On the bright side, if we accept the simple model in \S 3 by faith, 
  we can explain the basic features 
  in the log~N($>$S) -- log~S curves, 
  and also use the observed populations of X-ray sources 
  to probe the dynamical history of their host galaxy in the near past.      

\section{Summary} 

We have constructed a simple birth-death model 
  and calculated the luminosity function 
  of X-ray binaries in external galaxies. 
The model reproduces the features, such as the luminosity break, 
  in the log~N($>$S) -- log~S curves of spiral galaxies. 
By assuming a continual formation process,  
  the model also explains the absence of a luminosity break 
  for the disk sources in M81. 
The location of the luminosity break depends on the look-back time 
  of the previous starburst/star-formation activity, 
  and so it can be used 
  to probe the dynamic history of the host environment 
  of the X-ray sources. 
 
\section*{Acknowledgements}   

I thank Prof.~Don Melrose for generously providing me 
  with the research support in the last eight years. 
This work summarises a talk 
  presented at the Melrose Festschrift  
  in honour of his 60th birthday.  
The ideas presented in this work were developed 
  during the discussions with 
  Drs Richard Hunstead, Allyn Tennant, Douglas Swartz, 
  Helen Johnston, Kajal Ghosh and Roberto Soria, 
  and credits to the author, if any, 
  should also belong to all of them.     
I thank Dr Alex Finoguenov  
  for correspondence and for showing me 
  the luminosity function of the {\it Chandra} sources in M84 
  before publication, 
  Dr Roberto Soria for the {\it XMM-Newton} sources in M31, 
  and Dr Kajal Ghosh for the {\it ROSAT} image of M81 and M82 (Fig.~1).   
I would also thank Drs Martin Weisskopf and Allyn Tennant 
  for funding my visits to NASA-MSFC, where this work was started.    
This work is partially supported 
  by an ARC Australian Research Fellowship 
  and by a University of Sydney Sesqui R\&D Grant.  
   
\section*{Reference}  

\reference 
Blanton, E. L., Sarazin, C. L., Irwin, J. A. 2001, ApJ, 552, 106    
\reference 
de Loore, C. W. H., Doom, C. 1992, Structure and Evolution 
  of Single and Binary Stars (Dorchect: Kluwers   
  \indent Academic Press)   
\reference 
Fabbiano, G. 1995, in X-ray Binaries, eds. W. H. G. Lewin, 
  J. van Paradijs and E. P. J. van den  Heuvel  \indent
  (Cambridge: Cambridge University Press), 390  
\reference 
Fabbiano, G., Zezas, A., Murray, S. S. 2001, ApJ, 554, 1035 
\reference 
Finoguenov, A., Jones, C. 2001, ApJ, 547, L107  
\reference 
Garcia, M. R., et al. 2000, ApJ, 537, L23   
\reference 
Garcia, M. R., et al. 2001, 
  in Galaxies at the Highest Angular Resolution, 
  ASP Conf. Ser., in press (astro-ph/ \indent 0012387)  
\reference 
Irwin, J. A., Bregman, J. N., Sarazin, C. L. 2001, preprint     
\reference   
Johnston, H. M., Verbunt, F. 1996, A\&A, 312, 80  
\reference 
Kraft, R. P., et al. 2000, ApJ, 531, L9 
\reference 
Landau, L. D., Lifshitz, E. M. 1971, 
  The Classical Theory of Field (Oxford: Pergamon Press) 
\reference 
Matsumoto, H., et al. 2001, ApJ, 547, L25  
\reference 
Paczynski, B. 1971, Acta Astron., 17, 287   
\reference 
Pence, W. D., Snowden, S. L., Mukai, K., Kuntz, K. D. 2001, preprint 
  (astro-ph/0107133)
\reference 
Roberts, T. R., Warwick, R. S. 2000, MNRAS, 315, 98 
\reference 
Sambruna, R. M., et al, 2001, ApJ, 546, L9  
\reference 
Sarazin, C. L., Irwin, J. A., Bregman, J. N. 2000, ApJ, 544, L101    
\reference 
Schechter, P. 1976, ApJ, 203, 297   
\reference 
Shirey, R., et al. 2001, A\&A, 365, L195  
\reference 
Skumanich, A. 1972, ApJ, 171, 565  
\reference 
Supper, R., et al. 1997, A\&A, 317, 328
\reference 
Tennant, A. F., Wu, K., Ghosh, K. K., Kolodziejczak, J. J.,  
  Swartz, D. A. 2001, ApJ, 549, L43 
\reference 
Verbunt, F., Zwaan, C., 1981, A\&A, 100, L7   
\reference 
Wang, Q. D., Immler, S., Pietsch, W. 1999, ApJ, 523, 121 
\reference 
Weisskopf, M. C., Tananbaum, H. D., Van Speybroeck, L. P., 
  O'Dell, S. L. 2000, Proc. SPIE, 4012, 2  
\reference 
Wu, K. 1997, in `Accretion Phenomena and Related Outflows', 
  eds. D. T. Wickramasinghe, G. V. Bicknell \indent and L. Ferrario,  
  ASP Conf. Ser., 121 , 283  
\reference 
Wu, K., Tennant, A. F., Swartz, D. A., Ghosh, K. K., Hunstead, R. W. 
  2001, ApJ, submitted  
 
\end{document}